# High Mass Accuracy and High Mass Resolving Power FT-ICR Secondary Ion Mass Spectrometry for Biological Tissue Imaging

Donald F. Smith[1], Andras Kiss[1], Franklin E. Leach III[2], Errol W. Robinson[2], Ljiljana Paša-Tolić[2], Ron M.A. Heeren[1*]

1) FOM Institute AMOLF, Science Park 104, 1098 XG Amsterdam, The Netherlands
2) Environmental Molecular Sciences Laboratory, Pacific Northwest National Laboratory, Richland, WA 99352
* To whom correspondence should be addressed. E-mail: heeren@amolf.nl (f) +31-20-754-7100

**Abstract**

Biological tissue imaging by secondary ion mass spectrometry has seen rapid development with the commercial availability of polyatomic primary ion sources. Endogenous lipids and other small bio-molecules can now be routinely mapped on the sub-micrometer scale. Such experiments are typically performed on time-of-flight mass spectrometers for high sensitivity and high repetition rate imaging. However, such mass analyzers lack the mass resolving power to ensure separation of isobaric ions and the mass accuracy for elemental formula assignment based on exact mass measurement. We have recently reported a secondary ion mass spectrometer with the combination of a $C_{60}$ primary ion gun with a Fourier transform ion cyclotron resonance mass spectrometer (FT-ICR MS) for high mass resolving power, high mass measurement accuracy and tandem mass spectrometry capabilities. In this work, high specificity and high sensitivity secondary ion FT-ICR MS was applied to chemical imaging of biological tissue. An entire rat brain tissue was measured with 150 μm spatial resolution (75 μm primary ion spot size) with mass resolving power ($m/\Delta m_{50\%}$) of 67,500 (at $m/z$ 750) and root-mean-square measurement accuracy less than two parts-per-million for intact phospholipids, small molecules and fragments. For the first time, ultra-high mass resolving power SIMS has been demonstrated, with $m/\Delta m_{50\%} > 3,000,000$. Higher spatial resolution capabilities of the platform were tested at a spatial resolution of 20 μm. The results represent order of magnitude improvements in mass resolving power and mass measurement accuracy for SIMS imaging and the promise of the platform for ultra-high mass resolving power and high spatial resolution imaging.

**Introduction**

Secondary ion mass spectrometry (SIMS) has long been used for high spatial resolution chemical imaging of surfaces[1-3]. While matrix assisted laser desorption ionization (MALDI) uses a focused laser beam for desorption/ionization, SIMS uses a high energy primary ion beam, which can easily be focused to 1 µm or less for high spatial resolution imaging. While many applications lean towards spatial mapping of elements and small molecules, the use of SIMS for biological tissue and cell analysis has been steadily on the rise[4-7]. This is largely due to the advent of ion sources which produce large primary ions (e.g. bismuth[8]/gold[9,10] liquid metal ion guns and $C_{60}$ (buckminsterfullerene)[11] sources), which have higher secondary ion yields and induce less fragmentation than smaller primary ions[12]. The commercial availability of such sources has led to widespread use of SIMS for mapping of intact lipids and other small biological molecules at high spatial resolution[13-15].

Time-of-flight (TOF) mass spectrometers are most commonly coupled with SIMS sources. They offer high repetition rates and have high sensitivity, which are required for large (100's of µm) area imaging at high spatial resolution. Further, TOF instruments used for SIMS have moderate mass resolving power (typically $m/\Delta m_{50\%}$ = 2,000-10,000) and good mass measurement accuracy (10-100 parts-per-million; ppm). However, these types of instruments lack the mass resolving power to separate very closely based isobaric ions. Thus, selected ion images could be a convolution of multiple ions and may not represent the correct spatial localization of the species of interest. Further, the mass measurement accuracy (mma) is insufficient for exact mass molecular formula assignment, where a ppm error of ~1 is needed.

Fourier transform ion cyclotron resonance mass spectrometry (FT-ICR MS) offers both high mass resolution for resolving closely spaced ions, as well as high mma for confident molecular formula assignment. For FT-ICR MS, in-cell SIMS has been used with great success[16-19]. For MS imaging, the most straight-forward approach is to have an externally mounted primary ion gun for external ion accumulation before transfer to the FT-ICR cell for analysis. A number of groups have used externally mounted SIMS primary ion guns for high mass resolution secondary ion FT-ICR MS[20-23]. Further, the Winograd group has developed a commercial quadrupole orthogonal TOF MS with a $C_{60}$ primary ion gun for SIMS imaging at ~30 µm with mass resolving power ~14,000 and mma in the low ppm range[24-26]. The Vickerman group has also developed a $C_{60}$ buncher TOF instrument that provides high spatial resolution imaging (1 µm) with mass resolving power ~6,000 (at $m/z$ 500) [5,27,28]. These "hybrid" instruments both have the additional analytical capability of tandem MS for detailed analysis of secondary ions.

We have recently published the first combination of a polyatomic primary ion source ($C_{60}$) with an FT-ICR MS[29]. This system illustrates the additional benefit of high resolution mass spectrometry for SIMS imaging analysis, namely high mass accuracy (< 1 ppm), high mass resolving power ($m/\Delta m_{50\%}$ > 100,000) and MS/MS capabilities. In this manuscript we report improved MS imaging of biological tissue (here, rat brain), via improvements in the ion transfer optics. Current experimental figures of merit including mass measurement accuracy, mass resolving power and MS imaging spatial resolution are presented and discussed.

**Experimental**

$C_{60}$ Secondary Ion FT-ICR MS

$C_{60}$ secondary ion FT-ICR MS experiments were performed with a home-built system mated to a 9.0 T solariX FT-ICR MS (Bruker Daltonics, Billerica, MA, USA) which has been described previously[29]. Briefly, the instrument features a 40 keV $C_{60}$ primary ion gun (Ionoptika, Chandlers Ford, Hampshire, U.K.) operated in direct-current mode (DC) for fast production of secondary ion populations[26,28]. All experiments described in this paper used a primary ion beam energy of 20 keV and were done in the positive-ion mode. Here, the ion optics have been modified from the original $C_{60}$ SIMS FT-ICR MS design for improved sensitivity. Secondary ions are collected by an octopole ion guide and are then injected into one long quadrupole ion guide into the first multipole of the commercial solariX FT-ICR MS. This modification has eliminated one quadrupole and a conductance limit. Thus, the pressure in the sample chamber is higher than previously reported; $4\times10^{-6}$ mbar versus $2\times10^{-7}$ mbar. There is no observed rise of pressure in the $C_{60}$ primary ion source and no deleterious effects to the primary ion beam or secondary ion collection have been observed at this slightly higher pressure[24]. A schematic of the modified $C_{60}$ source can be found in electronic supplementary material Figure S1.

**Fig. S1** Schematic representation of the $C_{60}$ secondary ion source that has been coupled to a FT-ICR MS. The ion transfer optics were modified to include only an octopole for secondary ion collection and a quadrupole for transport of the ions to the commercial instrument's ion optics.

## FT-ICR MS Data Analysis

External mass calibration was performed using known indium oxide clusters from the ITO coated glass slide. Data was processed with the AMOLF developed Chameleon software[30]. Transients were apodized with a exponential function, zero filled once, fast Fourier transformed[31] and mass calibrated. Root-mean-square errors (rms) calculations, total ion current plots and ppm error plots were performed with in-house developed Matlab code (MATLAB version 7.13.0.564 (64 bit), Mathworks, Natick, USA).

## FT-ICR MS Rat Brain Tissue Imaging

Rat brain (Adult male Sprague-Dawley; Harlan Laboratories Inc., Livermore, CA, USA) was sectioned to ~12 µm on a cryo-microtome and thaw mounted on an ITO coated glass slide and stored at -20 °C until use. Profile spectra (Fig. 1) used a random stage raster during 3 s of secondary ion accumulation with a transient size of 1 megaword (MW). In the imaging mode, a surface exposure time (ion accumulation time) of 0.4 s per pixel was used with a primary ion beam spot size of 75 µm (primary ion dose of $1.2 \times 10^{12}$ ions/cm$^2$), a stage step size of 150 µm and transients of 2 MW were recorded. A calibration function that corrects for total ion abundance and relative ion abundance fluctuations was used[32,33,30]. Typically, hybrid ion trap/FT mass spectrometers will use a method to keep the number of ions injected into the FTMS as similar as possible (e.g automatic gain control), as to limit such mass shifts. However, such methods are not highly applicable to MS imaging, as the observed ion abundances can be erroneous due to different ion accumulation times depending on the total ion abundance. Thirteen mass spectra that span the range of total ion abundance observed over the rat brain section were used to calculate additional calibration terms for the total ion and relative ion abundance correction (peaks used were those of ITO clusters from the underlying ITO slide). In this way, ion-number fluctuations in the ICR cell, and thus mass fluctuations, due to tissue heterogeneity are corrected for. Mass spectra were apex peak picked at a signal-to-noise ratio (S/N) ≥ 5 and a "Mosaic Datacube" was generated with a mass bin size of 0.001 Da[30]. Exact mass assignments were made at mass accuracy tolerance of 2 ppm with possible elemental formulae $C_{0-100}H_{1-100}N_{0-3}O_{0-15}S_{0-2}P_{0-1}Na_{0-1}K_{0-1}$ in the MIDAS Molecular Formula Calculator v1.1 from the ICR Program at the National High Magnetic Field Laboratory. The LIPID MAPS database was used for assignment of intact phospholipids, with a search tolerance of 0.005 Da (LIPID Metabolites and Pathways Strategy; http://www.lipidmaps.org).

TOF-SIMS

The same rat brain section was analyzed on a Triple Focusing Time of Flight mass spectrometer (TRIFT II, Physical Electronics Inc., Chanhassen, MN, USA) with a gold liquid metal ion gun. Experiments were performed in static SIMS mode with 22 keV $Au^+$ primary ions in the positive-ion mode. The entire brain was analyzed in the mosaic mode (128 x 128 tiles with 256 x 256 pixels per tile), a raster size of 100 µm (pixel size = 391 nm) and 3 s acquisition per tile. Raw spectra were converted to the AMOLF "Datacube" structure with a mass bin size of 0.01 Da.

Transmission Electron Microscopy Grid

A transmission electron microscopy (TEM) grid (pitch = 83 µm, bar width = 10 µm; Agar Scientific, Stansted, Essex, U.K.) was applied on top of brilliant green dye (Staedler Lumocolor 318-5 permanent marker, Staedtler Mars GmbH & Co. KG, Nuernberg, Germany) to test the spatial resolution capabilities of the platform. Short transients (256 kilo-word) were collected for fast image generation. A primary ion beam spot size of 20 µm was used with a surface exposure time (ion accumulation time) of 0.75 s per pixel for a primary ion dose of $2.24 \times 10^{13}$ ions/cm$^2$.

**Results and Discussion**

Here, we report new developments on a $C_{60}$ secondary ion FT-ICR mass spectrometer for improved sensitivity and thus improved MS imaging on biological tissue. The prototype instrument had limited sensitivity, which resulted in the use of high primary ion doses (long accumulation time) in order to detect high quality mass spectra. As discussed above, the instrumental modifications entail the removal of one stage of differential pumping (a conductance limit and a quadrupole) and the extension of the remaining transfer quadrupole. These modifications have improved the sensitivity considerably, as illustrated by the lower primary ion dose needed for high quality spectra from biological tissue sections (reduced from $4.38 \times 10^{14}$ ions/cm$^2$ to $1.2 \times 10^{12}$ ions/cm$^2$, i.e. below the SIMS static limit). Further, the modifications have led to the detection of intact lipid species from biological tissue sections with good signal-to-noise ratio.

**Figure 1** shows a single mass spectrum from a rat brain section, zoomed into the lipid region of the spectrum.

Diacylglycerophosphocholine (PC) lipids are detected with good S/N (PC 32:0 ([M+K]$^+$) at *m/z* 772.5257 has a S/N of 68:1) and results in a rich spectrum in the lipid region. The abundant PC lipids are measured with exceptional mass accuracy, all within 1 ppm of the exact mass. The abundant peaks at *m/z* 713.4517 and 739.4835 are not intact lipids, rather the loss of trimethylamine (-N(CH$_3$)$_3$) from *m/z* 772.5257 and 798.5410, respectively. This fragmentation mechanism is associated with cation-lipid adducts (i.e. Na and K) and has been observed previously by C$_{60}$ SIMS and MALDI[34]. Fig. 1 exemplifies the advantage of coupling the polyatomic primary ion source with a high performance mass analyzer, specifically the matrix-free exact mass analysis of intact lipids from biological tissue sections.

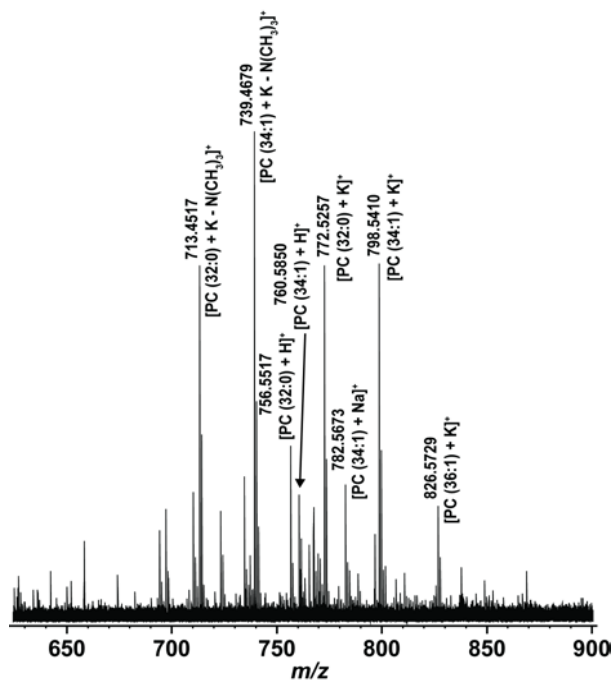

**Fig. 1** Zoom mass spectrum of the lipid mass range from a single C$_{60}$ secondary ion FT-ICR acquisition on rat brain tissue. Abundant diacylglycerophosphocholine lipids are detected with good signal-to-noise ratio and mass measurement accuracy.

While Fig. 1 illustrates the advantages of this platform for profiling analysis, the aforementioned instrument improvements have also improved the MS imaging capabilities. **Figure 2** contains various selected ion images from an MS imaging run of a rat brain section. A raster size of 150 µm was used in order to analyze the entire brain section in a reasonable amount of time (6,738 pixels, 17.5 hour analysis). However, the C$_{60}$ primary ion beam diameter was 75 µm, thus higher spatial resolution tissue imaging is possible. Fig. 2 illustrates the higher chemical diversity accessible with the improved ion transfer optics. At the bottom of Fig. 2 are selected ion images of intact PC lipids. It should be noted that the accumulation time per pixel in Fig. 2 was only 0.4 s, while that of Fig. 1 used a 3 s accumulation time and that previous experiments required up to 15 s for high quality spectra. Images are shown with a *m/z* bin size of 0.002 Da to ensure ultimate chemical specificity and the mass indicated is the average mass of all peaks detected above S/N = 5.

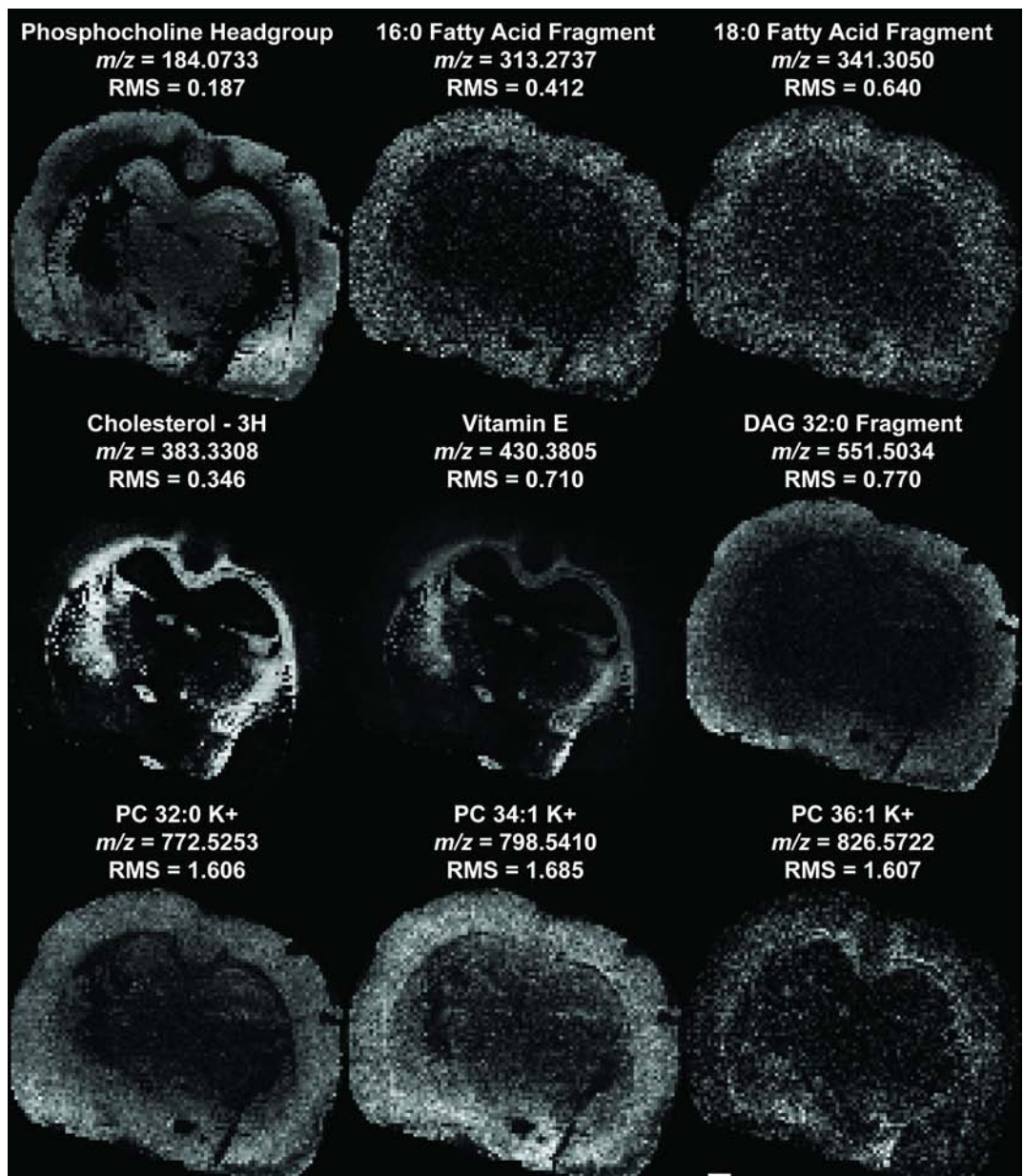

**Fig. 2** Ion selected images from a $C_{60}$ secondary ion FT-ICR MS imaging analysis of a rat brain tissue section. The identity of the ion is indicated, as well as the average mass of all peaks at that mass with S/N > 5. Root-mean-square error after abundance corrected mass calibration. Primary ion beam diameter was 75 µm, stage step size was 150 µm and the scale bar is 2 mm.

The data shown in Fig. 2 was calibrated with a calibration equation that corrects for total and relative ion abundance fluctuations, which cause frequency shifts and thus mass shifts (see Experimental section). The root-mean-square error for all ions shown in Fig. 2 was below 2 ppm. The rms error is higher at higher mass because the abundance correction parameters were calculated with respect to ITO clusters, which have the highest abundance below *m/z* 400. An interesting way to observe the results of the abundance correction calibration is to plot the ppm error of a selected ion *in each pixel* to generate a ppm error image. This is shown in **Figure 3**, along with the total ion count image and the ion selected image of cholesterol [M-OH+H]$^+$. **Fig. 3a** shows a total ion count image, where the highest abundance pixels are from the ITO slide surrounding the tissue section. On the brain tissue section, the corpus callosum shows a higher abundance than the surrounding tissue. Similarly, the cholesterol fragment at *m/z* 369.35 ([M-OH+H]$^+$) shows a high abundance in the corpus callosum (**Fig. 3b**). However, when the ppm error (external calibration) of the cholesterol fragment is plotted as an image, as in **Fig. 3c,** it can be seen that cholesterol is distributed throughout the tissue section, as expected. Further, the mass measurement accuracy within the corpus callosum is better than that of the surrounding tissue, which results in the white shading of the corpus callosum (ppm error closer to 0) and the blue shading of the surrounding tissue in Fig. 3c. This also indicates that the number of ions used for external calibration more closely matches the higher ion load from the corpus callosum than that from the surrounding tissue. However, when the abundance correction calibration is applied, as shown in **Fig. 3d,** the ppm error plot of the cholesterol fragment loses contrast, which shows that the calibration method corrects for the differences in ion abundance per pixel. Further, the mass accuracy is improved (centered near zero) thus the nearly uniform white image of Fig. 3d. Such high mass accuracy (and precision) ensures that ion selected images represent the analyte distribution and are not affected by ion number fluctuation due to tissue heterogeneity.

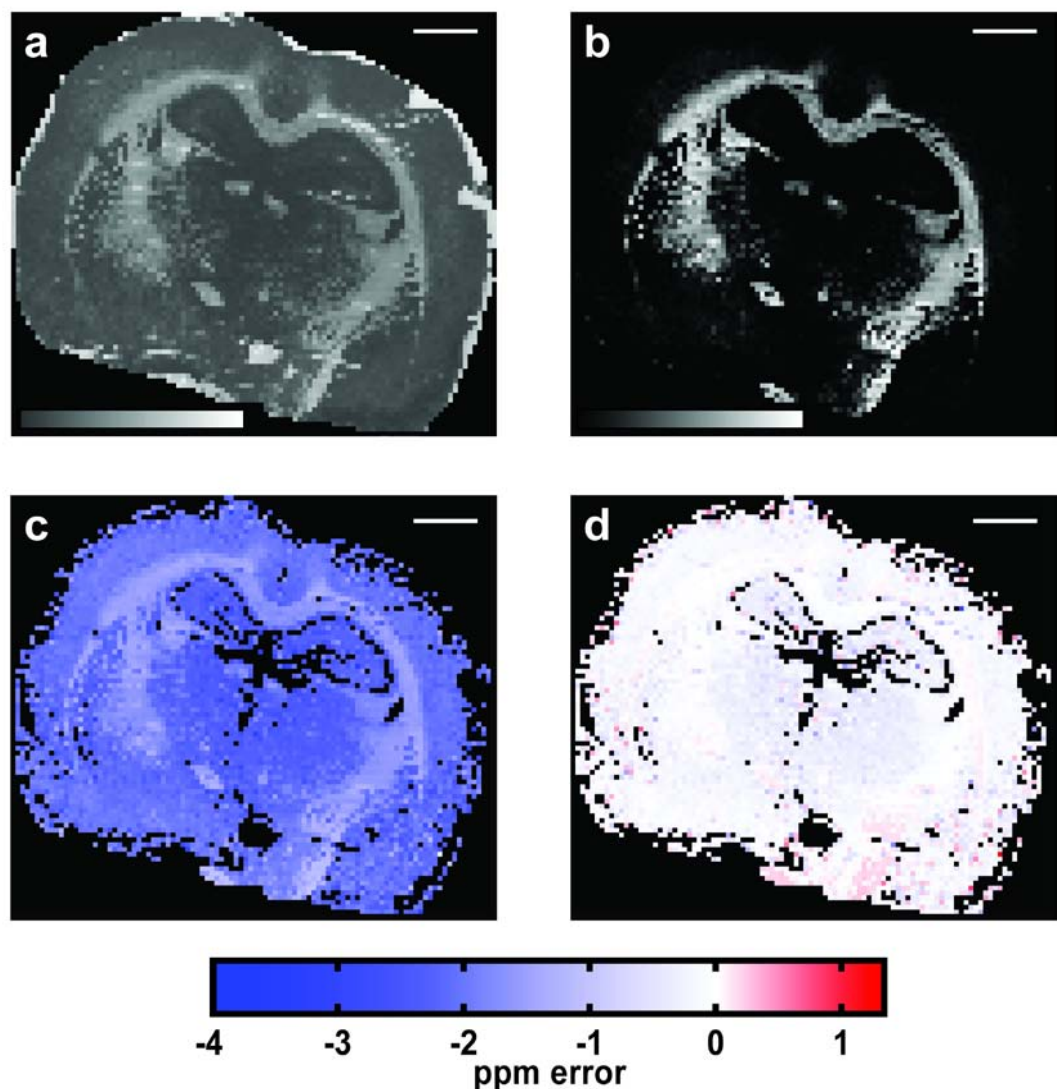

**Fig. 3** Results of correction of ion abundance fluctuations. a) total ion count image, b) ion selected image of cholesterol fragment (m/z 369.352 ± 0.001), c) ppm error plot for externally calibrated dataset and d) ppm error plot after abundance correction calibration. The scale bar is 2 mm.

The rat brain section shown in Figures 2 and 3 was also measured by TOF-SIMS. **Figure 4** shows an overlay of the summed mass spectra from FT-ICR MS and TOF-MS. The FT-ICR MS spectrum was binned at 0.001 Da and the TOF-MS spectrum at 0.01 Da (binning of the TOF-MS spectrum at 0.001 Da resulted in a discontinuous spectrum). It is clear from Fig. 4 that the FT-ICR provides more spectral detail than the TOF-MS. The zoom inset shows *m/z* 371, where the FT-ICR resolves seven peaks that are unresolved in the TOF-MS spectrum. As Fletcher and Vickerman discuss[35], for TOF-SIMS there is often a compromise between high spatial resolution imaging and high mass resolution. The TOF-SIMS spectrum shown in Fig. 4 was collected under optimum imaging conditions; that is (relatively) long primary ion pulses (19 ns, unbunched) for a stable spot at each position and faster data acquisition to allow analysis of the *whole* tissue section.  Importantly, the high mass resolving power of FT-ICR MS improves the chemical specificity for secondary ion mass spectrometry imaging with high mass resolving power, which is commonly compromised in TOF-SIMS imaging.

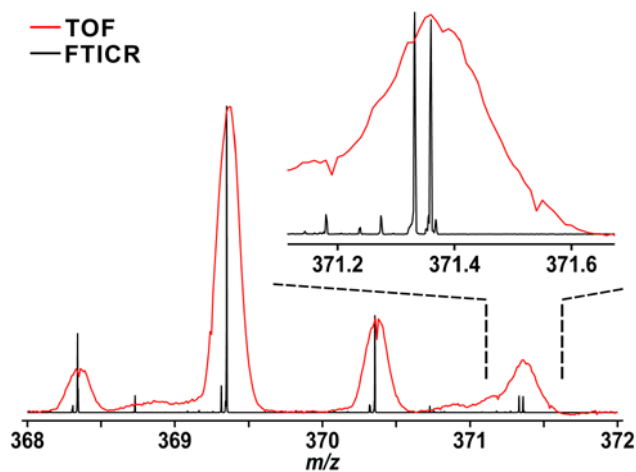

**Fig. 4** Spectral overlay of TOF-MS (red) and FT-ICR MS (black) of secondary ion mass spectrometry of the same rat brain section.  The high mass resolving power of the FT-ICR reveals mass features obscured in the TOF spectrum.  For comparison, the spectra were scaled to the highest abundance peak in each spectral window.

The spectra from Figures 2, 3 and 4 were collected with long transients for high mass resolving power over the entire mass range (2 MW transient data size).  At *m/z* 750, the maximum achievable mass resolving power ($m/\Delta m_{50\%}$) is 67,500 (126,500 at *m/z* 400).  However, in heterodyne mode ("narrow band") much higher mass resolving power is possibleby FT-ICR MS, as longer transients can be collected without extremely large transient data sizes, but at the cost of the mass range of interest.  The mass spectrum in **Figure 5** was collected in heterodyne mode in the mass region of the $In_2O_1$ cluster (*m/z* 245) from a single $C_{60}$ secondary ion FT-ICR MS analysis of an ITO coated glass slide.  In this way, the mass resolving power is extremely high, greater than 3,000,000 for all peaks.  Such ultra-high mass resolving power expands the capabilities of SIMS analysis to experiments such as sulfur counting[36,37] and isotopic fine structure analysis for exact mass assignment[36,38-40].

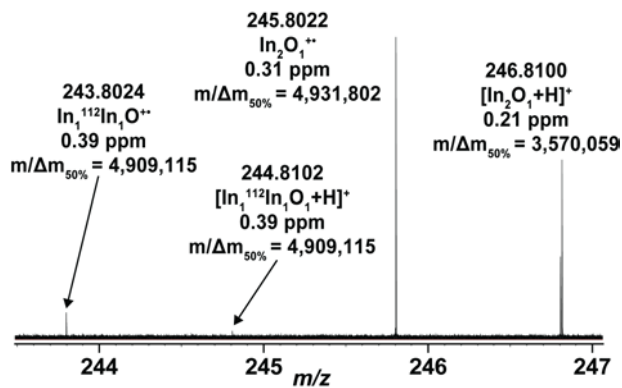

**Fig. 5** $C_{60}$ secondary ion FT-ICR mass spectrum of an ITO slide in heterodyne mode. The mass resolving power is extremely high, but at the expense of a narrow mass range.

The improvements in sensitivity described above also enable higher spatial resolution imaging, since a smaller primary ion beam probes a smaller surface area, thus a fewer number of secondary ions are generated per pixel. A TEM grid (pitch = 83 µm, bar width = 10 µm and hole = 73 µm) applied on top of brilliant green organic dye was used to test the spatial resolution capabilities of the platform, as shown in **Figure 6**. The ion selected image of the $M^{+\bullet}$ of brilliant green is shown (385.265 ± 0.005 Da, exact mass = 385.2638, RMS error = 1.47 ppm) and is detected from within the holes under the TEM grid. The locations of the grid bar are visible in the image, despite the probe size of 20 µm. This is due to the much lower abundance of the brilliant green when on top of a bar, even though some dye is still ionized by the slightly larger beam. In addition, the repetitive pattern of the grid is reproduced reliably and homogeneously, which owns to the stability of the primary ion gun and the stage positioning. While still a standard sample, these results are very promising for high spatial resolution $C_{60}$ secondary ion FT-ICR MS imaging.

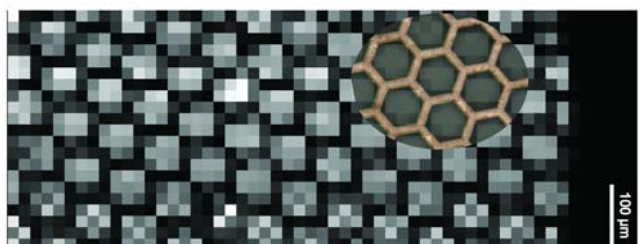

**Fig. 6** Ion selected image of brilliant green (385.265 ± 0.005 Da) from beneath a TEM grid (pitch = 83 µm, bar width = 10 µm and hole = 73 µm). The repetitive structure of the grid is reproduced well and the locations of the grid bars can be seen, even though the probe size is slightly larger than these features. An overlay with the optical image of the TEM grid is shown in the top right.

**Conclusions**

Here, we present high chemical specificity secondary ion MS imaging by Fourier transform ion cyclotron resonance mass spectrometry. Refinement of the ion transfer optics has improved the sensitivity of the platform and thus the quality of mass spectrometry imaging experiments. Primary ion doses below the static limit can now be used for high quality mass spectra. Further optimization of the ion transfer optics should provide additional sensitivity gains, for smaller primary ion beam sizes and better spatial resolution imaging.

The platform offers unprecedented mass resolving power and mass measurement accuracy for secondary ion mass spectrometry. Such high performance can provide exact mass identification of secondary ions with high specificity due to the high mass resolving power. Tissue imaging results in very rich mass spectra and can be used for complementary analysis with lower performance "conventional" time-of-flight secondary ion mass spectrometers which provide higher spatial resolution capabilities. The ultra-high mass resolving power capabilities of this platform can provide unique mass spectral analysis not possible elsewhere.

**Acknowledgements**

This work is part of the research program of the Foundation for Fundamental Research on Matter (FOM), which is part of The Netherlands Organization for Scientific Research (NWO). This publication was supported by the Dutch national program COMMIT. Portions of this research were supported by the American Reinvestment and Recovery Act of 2009 and the U.S. Department of Energy (DOE) Office of Biological and Environmental Research. The research described in this article was performed at the W. R. Wiley Environmental Molecular Sciences Laboratory (EMSL), a national scientific user facility sponsored by the Department of Energy's Office of Biological and Environmental Research and located at Pacific Northwest National Laboratory (PNNL). PNNL is operated by Battelle for the U.S. Department of Energy under Contract DE-AC05-76RLO 1830. D.F.S. would like to acknowledge the Alternate Sponsored Fellowship program at PNNL and R.M.A.H. the EMSL Wiley Visiting Scientist Fellowship program for support of portions of this work. We thank Jordan Smith and Chuck Timchalk (PNNL) for supplying the rat brain section and Julia Jungmann (AMOLF) for preparing the TEM grid sample.